\documentstyle[12pt]{article}

\newcommand{\be}{\begin{equation}}
\newcommand{\ee}{\end{equation}}
\newcommand{\bea}{\begin{eqnarray}}
\newcommand{\eea}{\end{eqnarray}}
\newcommand{\nn}{\nonumber \\}
\newcommand{\p}[1]{(\ref{#1})}
\newcommand{\ba}{\begin{array}}
\newcommand{\ea}{\end{array}}
\newcommand{\vs}[1]{\vspace{#1 mm}}

\newcommand{\e}{\epsilon}

\def\bbox{{\,\lower0.9pt\vbox{\hrule \hbox{\vrule height 0.2 cm
\hskip 0.2 cm \vrule height 0.2 cm}\hrule}\,}}
\newcommand{\dsl}{\pa \kern-0.5em /}

\newcommand{\pa}{\partial}

\def\One{{1\hskip -3pt {\rm l}}}
\def\II {\One}

\def\labell {\label}

\begin{document}

\topmargin 0pt
\oddsidemargin 5mm

\renewcommand{\thefootnote}{\fnsymbol{footnote}}
\begin{titlepage}

\setcounter{page}{0}

\rightline{\small hep-th/9909098}
\vskip -.75em\rightline{\small \hfill QMW-PH-99-13}

\vs{15}
\begin{center}
{\Large BPS States with Extra Supersymmetry}
\vs{10}

{\large
Jerome P. Gauntlett and Chris M. Hull\\}
\vs{5}
{\em   Department of Physics\\
       Queen Mary and Westfield College\\
       University of London\\
       Mile End Road\\
       London E1 4NS, UK}
\end{center}
\vs{7}
\centerline{{\bf Abstract}}
A state saturating a BPS bound  derived from a  supersymmetry algebra preserves
some fraction of the supersymmetry. This fraction of supersymmetry depends on
the charges carried by the system, and we show that in general there are
configurations of charges for which a BPS state would preserve more than half
the original supersymmetry. We investigate   configurations that could preserve
3/4 supersymmetry in string theory, M-theory and  supersymmetric field theories
and discuss whether states saturating these bounds actually occur in these
theories.
\end{titlepage}
\newpage
\renewcommand{\thefootnote}{\arabic{footnote}}
\setcounter{footnote}{0}

\section{Introduction}

It is well known that the supersymmetry algebra admits central
charges that give BPS  bounds on the energy. These charges
can be carried by solitons and when the bound is
saturated the states preserve some fraction of the supersymmetry. In addition
there are tensorial `central' charges carried by various $p$-branes
in string/M-theory, for example,
that lead to BPS bounds on the energy densities of the branes, and the BPS
$p$-branes preserve 1/2 of the supersymmetry \cite{az}. The
$p$-branes can intersect with or end on other branes while still preserving
some
supersymmetry,
and intersecting brane configurations have   been found that preserve fractions
$n/32$ of the supersymmetry for $n=0,1,2,3,4,5,6,8,16$
so that in each of these cases  no
more than half the supersymmetry is
preserved; see, for example, \cite{int, hyper, tseytlincomptwo,townohta,cal}.

By examining the supersymmetry algebra it is simple to see that
there must exist charges that would correspond to preservation of any
fraction $n/M$ of supersymmetry (where $M$ is the number of supersymmetries of
the system, so that $M=32$ for M-theory).
The  general anticommutator of $N$ supercharges $Q_{\alpha I} $ (with $\alpha $
a spinor index and $I=1,...,N$)
  can be written as
\be\label{susyalggen}
\{Q _A,Q _B\} = M_{AB}
\ee
where $A=1,...,M$ is a composite index $A=\{ \alpha I\}$
and $M_{AB}$ is a symmetric matrix of bosonic charges, which in most physical
systems will take the form
\be\label{mis}
  M_{AB}=H \delta _{AB} - Z_{AB}
\ee
with $H$ the hamiltonian and $Z_{AB}$ a traceless symmetric matrix of \lq
central'
charges which can be decomposed into a set of
$p$-form charges $Z^{IJ}_{\mu _1 ...\mu_p}$
contracted with gamma matrices.
Let the eigenvalues of $Z_{AB}$ be
 $\lambda _1,...,\lambda _M$ with
$\sum \lambda _A=0$. Then the supersymmetry algebra implies that  $M_{AB}$ must
be positive definite so that
the energy $E$ is bounded below by the largest eigenvalue, $E \ge \lambda $
where $\lambda = max \{ \lambda _1,...\lambda _M
\}$, as is easily seen in a basis
 in which $Z_{AB}$ is diagonal. If the  largest eigenvalue is $n$-fold
degenerate,
$\lambda _1=\lambda _2=...=\lambda _n \equiv\lambda$ say,
and if there is a state that saturates the bound with $E=\lambda$, then for
this state $M_{AB}$ will have $n$ zero eigenvalues
and by definition the state will preserve $n$ of the supersymmetries, namely
$Q_1,Q_2,...,Q_n$, and should fit into a supermultiplet
generated by the action of the remaining  $M-n$ supercharges.
Thus a  system will have a state  preserving a given fraction $n/M$ of
supersymmetry provided (i)
there is a configuration of charges such that the maximal eigenvalue of
$Z_{AB}$ is $n$-fold degenerate
and (ii) there is a state that saturates the BPS bound for these charges.

In many physical systems,  $Z_{AB}$ is an arbitrary symmetric traceless matrix,
since
a configuration  of charges   can be found that gives any desired  $Z_{AB}$.
For example,
in M-theory the matrix $M_{AB}$ has $32\times 33/2=528$ independent entries and
all 528 arise from the 11-momentum, a 2-form
charge and a 5-form charge, as $11$+$55$+$462$=$528$ \cite{DEM}. Moreover, each
of the
527 charges $Z_{AB}$ is believed to actually arise
in M-theory, and there is a 1/2-supersymmetric BPS state for each of the 527
charges \cite{CMH}. Most have been constructed explicitly,
while evidence for the occurrence of the M9-brane is given in \cite{BEH,MF}.
Then in M-theory there is a configuration of
charges
corresponding to each fraction $n/32$ of
supersymmetry for
$0 \le n \le 32$, and most can be realised without recourse to M9-branes.

If M-theory  is dimensionally reduced to one dimension by compactifying all the
spatial dimensions,
the resulting theory is a quantum mechanical   theory with 32 supersymmetries
and algebra (\ref{susyalggen}),(\ref{mis}), where $A=1,..., 32$ is now an
internal index
transforming under an $Sp(32)$ internal symmetry, and $Z_{AB}$ represents 527
scalar central charges,
transforming irreducibly under the $Sp(32)$  automorphism group of the
superalgebra,
which is a contraction of $OSp(32|1)$. All central charges are then clearly on
the same footing,
and there seems no reason why an arbitrary central charge matrix $Z_{AB}$, and
hence an arbitrary
fraction of supersymmetry $n/32$, should not be realisable.

If there is a set of charges in a supersymmetric theory
for which the maximal eigenvalue of  $Z_{AB}$ is
$n$-fold degenerate, and if there is  a state which saturates the  BPS bound,
it
would preserve $n/M$ of the supersymmetries.
In most cases that have been studied and for which the   state of lowest energy
has been found,
it turns out to  be a supersymmetric one saturating the bound.
The fact that most
allowed supersymmetric states actually occur suggests that it would be of
interest to investigate further the configurations
that could preserve exotic fractions of supersymmetry.

Our purpose here will be to give  some simple examples in which there is a BPS
bound for which a state  saturating it would
preserve 3/4 supersymmetry,
and to give some preliminary discussion as to whether such states actually
occur. The possibility of 3/4 supersymmetry has also been recently discussed in
\cite{Ueno},\cite{BL}.

We will first consider the  supersymmetry algebra in four dimensions.
It is straightforward to provide charges that lead to preservation of 3/4 of
the supersymmetry. This
algebraic structure can be embedded in higher dimensions and we will focus on
D=11. We will show that the charges preserving 3/4 of the supersymmetry
can be realised by considering a very simple configuration of
a membrane intersecting two fivebranes  according
to the array
\be
\matrix{
M5:&1&2&3&4&5& & & & & \cr
M5:&1& & & & &6&7&8&9& \cr
M2:&1& & & & && & & &\sharp \cr}
\labell{config}
\ee
where the symbol $\sharp$ is read as `10',
with the amount of supersymmetry preserved depending on the energy and the
charges of the three branes.
The case that has been discussed previously \cite{hyper, tseytlincomptwo} is
that in which
the product of all three brane charges is positive  (in our conventions),
leading to $1/4$ supersymmetry being preserved,
whereas we will find   new possibilities when   one of the branes  has negative
charge, and so is an anti-brane (or all three are anti-branes).
  We will analyse this case in some  detail and determine
under which conditions the fractions 1/4, 1/2 and 3/4 of the supersymmetry
could be preserved. One interesting feature is that for three or more
intersecting
branes,  switching all
the branes to anti-branes can lead to inequivalent results, whereas for
configurations with just two branes,
equivalent results would be obtained by the switch.
Many other configurations of branes with exotic supersymmetry in M-theory or
string theory can be generated from this
example by dualities.

\section{ Exotic Supersymmetry in D=4}

The general $N$ extended superalgebra in four dimensions is
\be
\{Q  ^I, \bar Q  ^J \} = -( P^\mu \delta ^{IJ}\Gamma_\mu+
V_\mu   ^{IJ}\Gamma^\mu+
iY_\mu ^{IJ}\Gamma^5 \Gamma^\mu
+X_{\mu\nu}^{IJ}
\Gamma^{\mu\nu}+iZ^{IJ}+i{\tilde Z}^{IJ}\Gamma^5)
\labell{algtwo}
\ee
where $Q^I$, $I=1,...,N$ is a Majorana spinor,   the charges $Y_\mu
^{IJ},Z^{IJ},{\tilde Z}^{IJ}$
are antisymmetric in the $IJ$ indices while $V_\mu   ^{IJ},X_{\mu\nu}^{IJ}$ are
symmetric and
$V_\mu   ^{IJ}$ is traceless.
In   supersymmetric theories, $P_\mu$ is the 4-momentum, $Z$ and $\tilde Z$ are
electric and
magnetic $0$-brane charges, $X_{\mu\nu}$ are domain wall charges \cite{AT},
$V_i,Y_i$ are string charges $(i=1,2,3)$ and $V_0,Y_0$  are charges for
space-filling 3-branes
\cite{CMH}. Moreover, in some cases $P_i$  could be a linear combination of the
 momentum and a
string charge, while $P^0$  could be a linear combination of the  energy and a
3-brane charge.

The number of charges on the right-hand-side of (\ref{algtwo})  is
$10\times N(N+1)/2+ 6\times N(N-1)/2$, which agrees with the number of
components,  $2N(4N+1)$, of
the  left-hand-side.
This suggests that by choosing the charges on the right hand side, it should be
possible to find a
system for which the BPS bound would
lead to any fraction $n/4N$ supersymmetry being preserved, provided that there
existed  a state
saturating the BPS bound.
In particular, there are some very simple systems that could allow $3/4$
supersymmetry.

For example, consider $N=2$ supersymmetry with   only the charges $P_\mu$ and
$Y_\mu ^{IJ}=Y_\mu
\e
^{IJ}$ non-zero.
A convenient choice  of gamma matrices is
\be
\Gamma^0= \pmatrix{
0 & i \cr
i & 0 \cr
}, \qquad
\Gamma^i= \pmatrix{
0 &i\sigma ^i \cr
-i \sigma^i & 0 \cr
}, \qquad
\Gamma^5= \pmatrix{
1 & 0 \cr
0 & -1 \cr
} \ee
Then configurations with $P^0=E$, $P^3=p$, $Y_0=u$ and $Y_3=v$ and all other
charges zero
have the superalgebra
\be
\{Q,Q^\dagger \} = diag(E-\lambda_1,E-\lambda_2,E-\lambda_3,E-\lambda_4)
\labell{algfo}
\ee
where $Q= (Q^1 +iQ^2)/\sqrt 2$ and the eigenvalues $\lambda _i$ are given by
\bea
\lambda_1&=&p+u+v\nn
\lambda_2&=&u-p-v\nn
\lambda_3&=&v-u-p  \nn
\lambda_4&=&   p-u-v
\labell{uvp}
\eea
Note that there is   a symmetry in the way the three charges occur.
Positivity implies that the energy $E$ satisfies $E\ge \lambda_i $ for each
$i$.
If only one of the charges is non-zero, $u$ say, then   $E \ge u$ and $E \ge
-u$
so that we obtain the standard bound $E\ge |u|$.
With two charges, $u$ and $v$ say, we obtain
$E \ge |u+v|$ and $E\ge |u-v|$ and when one of these is saturated
we have a  configuration
preserving 1/4 supersymmetry.
With all three charges, there are four bounds
corresponding to the four eigenvalues and in general when one is saturated
there will be  1/4 supersymmetry preserved.
However, for special values of the charges there can be degenerate eigenvalues.
Consider for example the case in which all charges are equal,  $u=v=p=-\lambda$
so that
\be
\{Q,Q\}=diag(H+3\lambda,H-\lambda,H-\lambda,H-\lambda)
\ee
If $\lambda$ is positive, a state with $E=\lambda$ would preserve the 3/4
supersymmetry corresponding to  supersymmetry
parameters of the form
$\epsilon=(0,\epsilon_2,\epsilon_3,\epsilon_4)$. For negative $\lambda$, a BPS
state with $E=-3\lambda$ would preserve 1/4
supersymmetry.

In \cite{GGHT}, it will be shown that a similar example occurs in the
Wess-Zumino model
with  $N=1$ supersymmetry. In that case, there is again a simple configuration,
corresponding to
intersecting domain walls with momentum along the intersection, for which a
state saturating the
bounds would have 1/4, 1/2 or 3/4 supersymmetry, depending on the values of the
charges. It will also be shown in \cite{GGHT} that the Wess-Zumino model
does not admit any classical configurations with 3/4 supersymmetry.

\section{Exotic Supersymmetry in String Theory and M-Theory}

\subsection{M-Theory}

The  general form of the eleven dimensional supersymmetry algebra has
\be
\{Q,Q\} =C (\Gamma^{M}P_{M} - {1\over 2!}\Gamma^{M_1 M_2}Z_{M_1 M_2} -
{1\over 5!}\Gamma^{M_1\dots M_5}Z_{M_1\dots M_5})  \ ,
\labell{susyalg}
\ee
where $C$ is the charge conjugation matrix, $P_{M}$ is the
energy-momentum 11-vector and $Z_{M_1 M_2}$ and $Z_{M_1\dots M_5}$ are
2-form and 5-form charges. 
The fraction of  supersymmetry that is preserved by
a configuration possessing
a given set of charges is given by the number of zero eigenvalues of the matrix
$\{Q,Q\}$ divided by 32. As argued in the introduction, both sides have equal
numbers of components (528), and all 528
charges on the right hand side actually arise in M-theory, provided we include
M9-branes carrying the charge
$Z_{0i}$
\cite{CMH}, so that there must be configurations of charges that could give
rise to all fractions $n/32$ of
preserved supersymmetry,  provided BPS states arise in that charge sector.

 For 3/4 supersymmetry, there  is a very simple set of
charges corresponding to two fivebranes and a membrane that allows
3/4 of the supersymmetry, obtained by embedding the example of the last section
in 11 dimensions and using dualities.
In addition we will show that there are
some novel combinations of three charges leading to the
preservation of 1/4 and 1/2
supersymmetry.
It is known \cite{hyper, tseytlincomptwo} that it is possible to have two
fivebranes and a membrane intersecting
according to (\ref{config}) and preserving 1/4 of the supersymmetry, provided
the product of all three brane charges is positive. Changing the signs of one
or three of the charges and tuning their values
allows 3/4 supersymmetry instead, as we shall see.

We begin by assuming that the only non-zero charges in \p{susyalg}
are
\bea
q_5 &=& Z_{12345}\nn
q_5' &=& Z_{16789\sharp}\nn
q_2 &=& Z_{1\sharp}
\eea
and positive charges will correspond to branes and negative charges to
anti-branes.
We use real gamma matrices with $C=\Gamma^0$ and $\Gamma^{0123456789\sharp}=1$.
It will be convenient to take a basis such that
\bea
\Gamma^{012345}&=&diag(1,1,-1,-1)\otimes \II _8  \nn
\Gamma^{016789}&=&diag(1,-1,1,-1)\otimes \II _8 \nn
\Gamma^{01\sharp}&=&diag(1,-1,-1,1)\otimes \II _8 \nn
\eea
where $\II _8 $ is
the $8\times 8$ identity matrix.
Setting $P^0=E$ we can then rewrite \p{susyalg} as
\be
\{Q,Q\} = diag(E-\lambda_1,E-\lambda_2,E-\lambda_3,E-\lambda_4) \otimes \II _8
\labell{algev}
\ee
where
\bea
\lambda_1&=&q_2+q_5+q_5'\nn
\lambda_2&=&-q_2+q_5-q_5'\nn
\lambda_3&=&-q_2-q_5+q_5'\nn
\lambda_4&=&q_2-q_5-q_5'
\labell{evs}
\eea
Since $\{Q,Q\}$ is a positive matrix we   have the BPS bound $E\ge  \lambda_i
$
for $i=1,2,3,4$.

 If there is only one non-zero charge,   $q_2$ say, then the BPS bound is
simply
$E\ge |q_2|$ and when it is saturated    1/2 of the supersymmetry is preserved.
For example, for BPS membranes (with $q_2$ positive and $E=q_2$)
we have
\be
\{Q,Q\} = diag(0,2q_2,2q_2,0)\otimes \II _8
\ee
The preserved
supersymmetry parameters
satisfy $\Gamma^{01\sharp}\e=\e$.

With an additional non-zero charge $q_5$, the BPS bounds are the two
conditions that $E\ge |q_2 +q_5|$ and $E\ge |q_2 -q_5|$. When either of
the bounds is saturated,  1/4 of the supersymmetry is   preserved. For example,
for a   membrane and a fivebrane,
\be
\{Q,Q\} = diag(0,2q_2,2(q_2+q_5),2q_5)\otimes \II _8
\ee
with 8 zero eigenvalues.
The supersymmetry preserved
is the intersection of that preserved by each of the membranes and fivebranes;
in this case
\be
\Gamma^{01\sharp}\e=\Gamma^{012345}\e=\e
\labell{susproj}
\ee

Adding the fivebrane to the membrane further halved the membrane's
supersymmetries to leave 1/4 supersymmetry.
However, a second fivebrane can now be added in the 16789 directions without
breaking any more supersymmetry,
as the corresponding projection $\Gamma^{016789}\e=\e$ on the supersymmetry
parameter
is already implied by the
conditions
(\ref{susproj}).
We can indeed add
a third positive charge $q_5'$ and preserve all 8 supersymmetries if
the energy saturates the BPS bound,  $E=q_2+q_5+q_5'$.
We will refer to this as the usual BPS intersection of
the $(2,5,5)$ system as it has been extensively studied in the literature.
An identical analysis goes through for
$(2,\bar 5,\bar 5')$ if we take $E=q_2-q_5-q_5'$,  for
$(\bar 2,5,\bar 5')$ if we take $E=-q_2+q_5-q_5'$ and for
$(\bar 2,\bar 5,5')$ if we take $E=-q_2-q_5+q_5'$; in each case, we can start
with any two of the branes intersecting and
preserving 1/4 supersymmetry, and then add the third for free without any
further breaking.

Returning to the $(2,5)$ system preserving 8 supersymmetries, adding
   an anti-fivebrane with $\Gamma^{016789}\e=-\e$ instead of a fivebrane to
give the $(  2,5,\bar 5')$ configuration
 would appear to break the original 8 supersymmetries of the
membrane-fivebrane system, but, as we shall show, the BPS bound leads to
8 supersymmetries  if the energy saturates the bound.
 (These will be a different    8-dimensional subset  of the 32
for some values of the charges and will be
the     same 8 for other values.)  The situation is the same for the $(\bar
2,\bar 5,\bar 5')$, $(\bar 2,5,5')$
and $( 2,\bar 5,5')$
  systems; in each case any two of the three branes preserve  8
supersymmetries, while the third brane appears to
  break  these 8 supersymmetries, but nonetheless 8 supersymmetries would be
preserved if the
bound is saturated.
Moreover, if such a 1/4 supersymmetric BPS state exists, tuning the charges to
particular values enhances the number of
supersymmetries to 16 or to the exotic value of 24.

For general charges it is useful to contrast the
analysis for  configurations related by switching branes with
anti-branes and we will focus on the $(2,5,5')$ and $(\bar 2,\bar 5,\bar 5')$
systems.
With this in mind we return to \p{algev} and \p{evs} and first consider
the $(2,5,5')$ case in which
all the charges are positive.  Clearly $\lambda _1= q_2+q_5+q_5'$
is the largest eigenvalue  and hence the BPS bound is
$E\ge q_2+q_5+q_5'$ and when it is saturated we preserve 1/4 of the
supersymmetry; this is the usual case considered above. To
analyse the $(\bar 2,\bar 5,\bar 5')$ case in which  all charges are negative,
it is useful to rewrite \p{evs} as
\bea
\lambda_1&=&q_2+q_5+q_5'\nn
\lambda_2&=&-(q_2+q_5+q_5')+2q_5\nn
\lambda_3&=&-(q_2+q_5+q_5')+2q_5'\nn
\lambda_4&=&-(q_2+q_5+q_5')+2q_2
\labell{evstwo}
\eea
One of $\lambda_2,\lambda_3,\lambda_4$ is now the biggest eigenvalue and is
positive, since $\lambda _1$ is negative and the sum of
the eigenvalues is zero. For example, when
\be
 0\ge q_5\ge q_5', \qquad 0\ge q_5\ge q_2
\labell{qcon}
\ee
it is
$\lambda_2$ that is the biggest and the BPS bound
is $E\ge-(q_2+q_5+q_5')+2q_5$. If this bound is saturated
\be
\{Q,Q\} = diag(-2(q_2+q_5'),0,2(q_5-q_5'),2(q_5-q_2)) \otimes \II _8
\ee
 and  1/4 of
the supersymmetry would be preserved.
To obtain exotic preservation of 1/2 or 3/4 supersymmetry we only
need to tune the charges:
1/2 supersymmetry is preserved when either $q_5=q_5'$ or $q_5=q_2$
and 3/4 supersymmetry is preserved when $q_5=q_5'=q_2$.

Thus, for the $(\bar 2,\bar 5,\bar 5')$ system with charges satisfying
\p{qcon}, the lowest energy allowed by supersymmetry
is  $E=-(q_2+q_5+q_5')+2q_5$. If the ground state of this system indeed has
this energy, then it would preserve 1/4
supersymmetry for generic values of the charges, but when two of the charges
are equal, 1/2 supersymmetry would be preserved
and if all three charges are equal, 3/4 supersymmetry would be preserved.
Similar results follow for the cases in which it is $\lambda _3$ or $\lambda
_4$  that is the
biggest.

To obtain further insight,
we continue with the case with charges satisfying \p{qcon}, so that
$\lambda_2$ is
the largest eigenvalue.
The specific 8 supersymmetries that would be preserved are the same as
those preserved by the two intersecting branes $(\bar 2,\bar 5')$. It is
interesting that it is
these two branes that are contributing to
the energy positively while the other fivebrane is contributing
negatively.
 Recall that if we add
a fivebrane,  with $q_5$ positive, to $(\bar 2,\bar 5')$ to obtain the
$(\bar 2,5,\bar 5')$ configuration, the usual case of preservation of 1/4
supersymmetry is
obtained if we have  $E=-(q_2+q_5')+q_5$. The new point here is that we can
add instead an anti-fivebrane, with $q_5$ negative, to get the $(\bar 2,\bar
5,\bar 5')$ system
and still preserve the same supersymmetry
if again $E=-(q_2+q_5')+q_5$, as long as $q_5\ge q_5'$ and $q_5\ge q_2$.
In either case, the naive energy would just be the sum of the
energies of the three branes, i.e. $E_n=|q_2|+|q_5|+|q'_5|$. This is the
correct
result for the
usual 1/4 supersymmetric
$(\bar 2,5,\bar 5')$  case, but for the exotic $(\bar 2,\bar 5,\bar 5')$ case
the energy of a state saturating the bound would be $E=E_n -  V$ where $V=
2|q_5|$, suggesting that $V$ might
be interpreted as
some kind of binding energy or as some tachyonic contribution.

Finally, note that if
the conditions \p{qcon} are not both satisfied, then either
$\lambda _3$ or $\lambda_4$ will be the largest eigenvalue and adding
the anti-fivebrane to $(\bar 2,\bar 5')$ will break the original 8
supersymmetries and  lead to a {\it different } 8
supersymmetries being preserved.

\subsection{Tachyon Condensation}

It is perhaps worth comparing the above with the case of coincident
brane/anti-brane
pairs.
It
has been argued that $m$ D-branes and $m$ anti-D-branes will completely
annihilate to leave the vacuum with   energy
$E=mT+mT-V=0$ where
$T$ is the energy of a single brane and the contribution
 $V=2mT$
arises from the negative potential energy released by tachyon condensation
\cite{sen170,sen207}.
  Duality then implies that this should also apply to any $m$ brane/anti-brane
pairs in M-theory or string theory, which
should again   completely annihilate.
The tachyon condensation reduces the energy to the minimum allowed by the BPS
bound, which in this case is zero as
the brane/anti-brane pair carries no net charge.
Adding a further $n$ branes to obtain $n+m$ branes and $m$ anti-branes, the $m$
anti-branes should completely
annihilate
$m$ of the branes to leave $n$ branes with energy $E=nT$, which can be written
as $E=E_n-V$ where
$E_n=(2m+n)T$ is the naive energy given by the sums of the energies of the
individual branes and anti-branes and $V=2mT$.

Then
for two coincident $p$-branes     of charges $q$,  $\tilde q $, the naive
energy of the system would  be the sum of the
energies of the respective branes, $E_n=|q|+|\tilde q |$.
This is the correct energy if $q,\tilde q $ have the same sign, so that they
are either both branes or both anti-branes.
However, if the charges have opposite sign so that one is a brane and the other
an anti-brane (e.g. $q=(n+m)T $ and
$\tilde q =-mT$ for the case above),
the resulting configuration has $E=|q+\tilde q | =E_n-V$
with $V=2 \, min(|q|,|\tilde q |)$.

This is suggestively similar to the case considered above when
$\lambda_2$ is the largest eigenvalue.
The $(\bar 2,\bar 5')$ system enters the energy formulae in exactly the same
way as a 5-brane of charge
$\tilde q_5=-(q_2+q'_5 )$ would. Adding a fivebrane with positive charge $q_5$
gives a system with $E=q_5+\tilde q_5$, while
adding an anti-fivebrane with negative charge $q_5=-q$ gives a system with $E=
\tilde q_5-q$ and $V=2q$. More generally, 
when the product of the brane
charges is positive, the naive energy is the sum of the energies
of the branes $E_n$=$|q_2| + |q_5| + |q_5'|$ and such configurations
preserve 1/4 of the supersymmetry. When the product of the charges
is negative, exotic preservation of supersymmetry is possible only if
the naive energy is modified to $E=E_n-2Min(|q_2|,|q_5|,|q_5'|)$.
This suggests that tachyon condensation could play a role here also, reducing
the energy below the sum of the brane energies.
It seems plausible that this could indeed
be the case and that it reduces the energy to
the minimum allowed by supersymmetry.

\subsection{String-Theory}

The M-theory example   with an M2-brane and two M5-branes is related by duality
to many other configurations of three branes in
string theory or M-theory.
For example, it is related to the   type II  configuration of a Dp-brane, a
D(8-p) brane and a fundamental string
intersecting in a point, with the  Dp-brane in the directions $1,2,...,p$, the
D(8-p) brane in the directions $p+1,p+2,...,8$
and the fundamental string in the 9th direction, or to the configuration of a
D5-brane in the 12345 directions, a
NS5-brane in the 12678 directions and a D3-brane in the 129 directions studied
by Hanany and Witten \cite{hananywitten}.
In each case there is the usual 1/4 supersymmetric configuration in which one
of the three branes is added \lq for free', and
an exotic configuration obtained from this by reversing the orientation of one
of the branes, or of all three branes,
in which a BPS state would preserve
  1/4 supersymmetry   for generic charges and 1/2 or 3/4 supersymmetry  when
two or three of the charges are of equal
magnitude. There
are no such configurations with only D-branes, so that the methods of
\cite{sen170,sen207} cannot directly be used to test the possibility of tachyon
condensation leading to   exotic BPS states.

\section{Conclusion}

We have shown that the supersymmetry algebra allows configurations preserving
exotic
amounts of supersymmetry and we have identified  simple
configurations of charges in M-theory and in field theories
such that  any state with
the
lowest energy allowed by supersymmetry  would preserve 3/4 supersymmetry, but
we have not been able to establish whether
or not such
states actually occur. We have been unable to find any D=11 supergravity
solutions with 24 Killing spinors, corresponding to 3/4 supersymmetry.
The known supersymmetric supergravity solutions with two (anti)-fivebranes 
and an (anti)-membrane either preserve 1/4 or none of the
supersymmetry \cite{hyper,tseytlincomptwo}. 
We take  the mass parameters associated with the harmonic or
generalised harmonic functions of each of the individual 
branes  to be positive. Then if the product of the three charges
is positive (for example, the  $(2,5,5')$ or 
$(2,\bar 5,\bar 5')$ configurations), 
then the solutions have 8 Killing spinors 
corresponding to 1/4 supersymmetry.
When the product of the charges is negative,
(for example, the $(\bar 2,\bar 5,\bar 5')$ configuration),  
the known supergravity solutions actually break all of the supersymmetry.
If we vary the signs of the mass parameters we do not obtain solutions
with more Killing spinors\footnote{This point was also discussed in 
\cite{popelu}.}. 
It is also possible to prove that no 3/4 supersymmetric classical 
solutions of the Wess-Zumino
model exist, even though supersymmetry would have allowed them
\cite{GGHT}.

In each of our examples, there are no spatial dimensions that are transverse to
all the branes and in such situations a
number of subtleties can arise, but nonetheless  our analysis does recover the
known cases of supersymmetric intersections.
The configurations we have identified are parameterised by three charges. For
generic values of these charges, a BPS state would preserve 1/4 supersymmetry,
but
for special values when two or three of the charges are equal, a BPS state
would preserve 1/2 or 3/4 supersymmetry, respectively.

We have conjectured that in string theory and M-theory, tachyon condensation
could play a role in reducing the energy to the
minimum allowed by the BPS bound, just as it does for the  brane/anti-brane
pair.

It would be very interesting to either establish that such exotic states do
occur in certain theories, or, if they don't
exist, to understand  the reason for this, given that supersymmetry appears to
allow them. We hope to return to these
issues in the future.

\vskip 1cm

\noindent{\bf Acknowledgements:}
We thank Gary Gibbons, Jeff Harvey, Michael Douglas, David Tong and
Paul Townsend for helpful discussions.

\end{document}